\documentclass[aps,pra,twocolumn,floatfix]{revtex4-2}

\usepackage{amsmath,amssymb,graphicx,bbold}
\usepackage{graphicx}  
\usepackage{dcolumn}   
\usepackage{bm}        
\usepackage{color}
\usepackage{hyperref}
\usepackage{xcolor}

\usepackage{xcolor}

\usepackage{physics}
\usepackage{ulem}
\usepackage{subfigure}

\begin{document}

\title{Gouy phase engineering of self-splitting quantum correlations}
	
	\author{A. L. S. Santos Junior$^1$, M. Damaceno$^2$, A. C. Barbosa$^1$, N. A. Ribeiro$^1$, G. B. Alves$^1$, \\ P. H. Souto Ribeiro$^2$ and A. Z. Khoury$^1$}
	\affiliation{1- Instituto de Física, Universidade Federal Fluminense, 24210-346 Niterói, RJ, Brazil}
	\affiliation{2- Departamento de Física, Universidade Federal de Santa Catarina, 88040-900 Florianópolis, SC, Brazil}
	\date{\today}
	
	\begin{abstract}
		In this work, we demonstrate the effect of self-splitting spatial quantum correlations induced by Gouy phase engineering. In the process of spontaneous parametric down conversion the pump beam is structured with a mode superposition that produces a dynamical splitting and recombination of the light beam. This structure is transferred to the quantum correlations between signal and idler photons. As a result the joint two-photon probability distribution propagates like a self-splitting and recombining light beam, implementing a Mach-Zehnder-like interferometer. We observe heralded single-photon interference and two-photon NOON state interference. These results open new avenues for applications in quantum metrology.
        \end{abstract}
	
	\maketitle

\label{sec:intro}

\textit{Introduction.} -- Quantum correlations between photon pairs generated via spontaneous parametric down-conversion (SPDC) have become an established and powerful tool for quantum technologies \cite{walborn2010}. Since its prediction by D.N. Klyshko in the 1960s \cite{zeldovich69} and experimental demonstration by Burnham and Weinberg in 1970 \cite{Burnham70}, several unique features have been revealed, such as Hong–Ou–Mandel interference \cite{hom87}. Nonlocal correlations in the transverse spatial degrees of freedom of SPDC photons were thoroughly investigated in Refs. \cite{ph94,strekalov95} and their quantum nature could be firmly established later through the violation of a classical inequality \cite{howell04}.

The mechanism responsible for these transverse spatial correlations, as well as their connection to the SPDC pump beam, was demonstrated in Ref. \cite{monken98}. Building upon this important result, dozens of applications were subsequently proposed and experimentally realized \cite{walborn2010}. Most of these studies explored the transfer of the angular spectrum of the pump laser into the spatial correlations of the photons, in order to observe interesting effects at fixed detection planes. In other words, the dynamical aspects of the propagation of these correlations were not investigated in the majority of works. An exception is the study demonstrating the generation of the equivalent of an Airy beam within the two-photon correlations \cite{santosh18,lib20}.

The experimental realization of structured light fields capable of exhibiting exquisite dynamical features, such as \textit{bottle beam} distribution and pattern revivals, through Gouy phase engineering was demonstrated in Refs. \cite{arlt2000,Braian2020}. 
Later, this idea was combined with polarization \cite{Bao-Sen2021} and time envelope \cite{Shen2025}, giving rise to revivals in multiple degrees of freedom. Engineering the Gouy phase has become a powerful tool for shaping both the transverse and longitudinal spatial distribution of paraxial light beams, which has been used for multiple purposes from particle trapping in a dark focus \cite{Melo2020,Almeida2023} to high-dimensional quantum protocols \cite{Mao2022}. More recently it has been proposed to protect light structure in random media \cite{Shen2026}. The role played by the Gouy phase in the generation of entangled photon pairs through spontaneous parametric down conversion has been discussed in connection with the two-photon wave function \cite{DEBRITO2021}. The improved phase sensitivity provided by NOON states \cite{Dowling2008,Giovannetti2011} has also been demonstrated for the Gouy phase using photon pairs generated by SPDC \cite{Hiekkamaki2022}. In this case, different Gouy phases were imprinted in each photon of the pair after exiting the nonlinear crystal and the NOON state was created by Hong-Ou-Mandel interference on a subsequent beam splitter.

In the present work, we investigate the transfer of the spatial structure and its propagation dynamics of a pump beam to quantum correlated photon pairs. This is achieved by engineering a pump beam that undergoes self-bifurcation and recombination in two dimensions, which we refer to as a self-splitting beam, through the superposition of spatial modes with different Gouy phases. We show that these pump-beam features can be imprinted onto the quantum correlations, giving rise to single- and two-photon nonlocal self-splitting beams. In the single-photon regime, we study the effect of an absorbing obstacle and show that Gouy-phase engineering helps preserve the relevant spatial and quantum features in an obstructed channel. In the two-photon regime, the resulting self-splitting beam can be interpreted as a NOON-like spatial state with N=2, accessed through pump-induced correlations. Since this dynamics emulates a Mach–Zehnder-like interferometer, we exploit it by introducing a phase shift with a glass plate, thereby demonstrating the super-resolution associated with the effective low-N NOON state and establishing a simple route towards metrological applications.

\textit{Theory.} -- The free propagation of a paraxial beam with waist $w_0$ and 
Rayleigh length $z_R = \pi w_0^2/\lambda$ along the \(z\)-axis can be described by Hermite-Gauss (HG) modes
{\begin{equation}
\begin{aligned}
\label{HGmodes}
%
HG_{m,n}(\mathbf{\tilde{r}})
&= \frac{C_{m,n}}{\sqrt{w(\tilde{z})}}\,H_m\!\left(\sqrt{2}\,\tilde{x}\right)H_n\!\left(\sqrt{2}\,\tilde{y}\right)
\\
&\times
e^{-(1+i\tilde{z})(\tilde{x}^2+\tilde{y}^2)}\,e^{-i\phi_N(z)}\,,
\end{aligned}
\end{equation}
where \(H_j\) denotes the Hermite polynomial of order \(j\), 
the normalized coordinates are $\mathbf{\tilde{r}}=(x/w,y/w,z/z_R)$ and
\(w(z) = w_0 \sqrt{1+\tilde{z}^2}\) is the beam width. The normalization constant is 
\(C_{m,n}=(2/\pi)^{1/2}/\sqrt{2^N m!\,n!}\)\,. The last term in Eq.~\eqref{HGmodes} is the Gouy phase, whose argument 
is given by
\begin{equation}
\phi_N(z)=(N+1)\arctan(\tilde{z})\,,
\end{equation}
and \(N=m+n\) denotes the total order of the HG mode. The Gouy phase plays a fundamental role in this analysis, as it depends not only on the propagation coordinate \(z\), but also on the mode indices \(m\) and \(n\), which determine the transverse structure of the field. Consequently, when the beam is prepared as a superposition of HG modes, each component with a different value of \(N\) acquires a distinct Gouy phase and therefore propagates with its own effective phase velocity. This establishes a direct interplay between the longitudinal evolution of the beam and its transverse spatial structure. Such a dual dependence is precisely what enables the Gouy phase to act as the central mechanism for engineering nontrivial three-dimensional field structures.

We now introduce a class of beams whose transverse intensity undergoes 
self-bifurcation upon free propagation. We refer to this field as a 
\textit{self-splitting beam}, defined by
\begin{equation}
\label{Psi}
    \Psi = HG_{0,0} - \sqrt{2}\, e^{i\theta_c}\, HG_{2,0}\,.
\end{equation}
Figure~\ref{fig:superposition} shows the resulting transverse intensity 
pattern as a function of the propagation distance \(z\) for two values 
of the control phase, \(\theta_c = 0\) and \(\pi\). In 
Fig.~\ref{fig:superposition}(a) (\(\theta_c=0\)), two lobes merge into 
a single central peak and then split again, in direct analogy with the 
input--output port structure of a beam splitter. Since the Gouy phase 
enters the superposition only as a relative phase between the two 
HG components, the entire longitudinal evolution can be 
reproduced, up to diffractive effects, on a fixed plane (\(z=0\)) 
by tuning the control phase \(\theta_c\). This one-to-one correspondence 
between \(z\) and \(\theta_c\) is indicated on the upper axis of 
Fig.~\ref{fig:superposition}(a). The complementary case \(\theta_c=\pi\), 
shown in Fig.~\ref{fig:superposition}(b), displays the reversed dynamics: 
a peak initially centered on the optical axis splits into two lobes and 
subsequently recombines, realizing a Gouy-phase-induced analog of a 
Mach--Zehnder interferometer. The same evolution is again recovered at 
the fixed plane \(z=5z_R\) by scanning \(\theta_c\).

\label{sec:theory}
\begin{figure}
    \centering
    \includegraphics[width=0.9\linewidth]{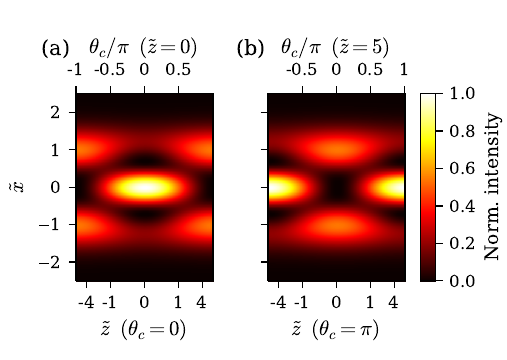}
    \caption{Self-splitting beam dynamics under free propagation. Transverse 
intensity of the superposition given by Eq.\eqref{Psi} as a function of \(\tilde{z}\) for (a) \(\theta_c=0\) and 
(b) \(\theta_c=\pi\). The upper axis shows the one-to-one mapping 
between \(z\) and the control phase \(\theta_c\) applied at 
\(\tilde{z}=0\) in (a) and at \(\tilde{z}=5\) in (b).}
    \label{fig:superposition}
\end{figure}

The transfer of the angular spectrum of the pump beam to the two-photon state generated by spontaneous parametric down conversion~\cite{monken98} constitutes a powerful tool, by which any engineered pump profile is inherited by the signal–idler pair, as detailed below. In the case of interest here, we apply this theory to imprint the self-splitting mode structure onto the correlations of the photon pair, enabling us to construct three-dimensional spatial dynamics in photon pairs. This mechanism thus provides a natural route to carry the Gouy-phase-induced dynamics from the classical to the quantum domain.

Following Refs. \cite{monken98} and \cite{walborn2010}, we can calculate the coincidence counting rate as a function of the pump beam angular spectrum or amplitude distribution and geometrical parameters of the system.  It is proportional to the fourth-order correlation function
\begin{equation}
C(\boldsymbol{\rho}_s, \boldsymbol{\rho}_i) = \|\mathbf{E}^{(+)}(\boldsymbol{\rho}_s) \mathbf{E}^{(+)}(\boldsymbol{\rho}_i) | \psi \rangle\|^2\,,
\end{equation}
where \(\mathbf{E}\) denotes the electric field operator, while \(\boldsymbol{\rho}_s\) and \(\boldsymbol{\rho}_i\) are the transverse coordinates of the signal and idler detectors at their respective detection planes. 

Since \(|\psi\rangle\) is a two-photon state, the correlation function takes the form
\begin{equation}
C(\boldsymbol{\rho}_s, \boldsymbol{\rho}_i) = |\Psi(\boldsymbol{\rho}_s, \boldsymbol{\rho}_i)|^2\,,
\end{equation}
which can be interpreted as the conditional two-photon detection probability, with
\begin{equation}
\Psi(\boldsymbol{\rho}_s, \boldsymbol{\rho}_i) = \langle \mbox{vac} | \mathbf{E}^{(+)}(\boldsymbol{\rho}_s) \mathbf{E}^{(+)}(\boldsymbol{\rho}_i) | \psi \rangle
\end{equation}
being the two-photon wavefunction. Within the paraxial approximation, the positive-frequency component of the electric field reads
\begin{equation}
\mathbf{E}^{(+)}(\boldsymbol{\rho}) = \mathcal{C}_1 \int \mathrm{d}\mathbf{q} \, \mathbf{a}(\mathbf{q}) 
\,e^{\left[i\left(\mathbf{q} \cdot \boldsymbol{\rho} + \sqrt{k^2 - q^2} \, z\right)\right]},
\end{equation}
\(\mathcal{C}_1\) is a constant, \(\mathbf{a}(\mathbf{q})\) is the annihilation operator in the transverse momentum space, 
\(\boldsymbol{\rho}\) represents the transverse spatial coordinate, and \(z\) is the longitudinal coordinate.

The coincidence count rate can be calculated using the SPDC two-photon state. In the thin crystal and paraxial approximation, it is given by \cite{walborn2010}: 
\begin{equation}
|\psi\rangle = \mathcal{C}_2 \iint_{D} d\mathbf{q}_s \, d\mathbf{q}_i \, v(\mathbf{q}_s + \mathbf{q}_i) 
|\mathbf{q}_s \rangle \, |\mathbf{q}_i \rangle ,
\end{equation}
where $\mathcal{C}_2$ is a constant and $v(\mathbf{q}_p = \mathbf{q}_s + \mathbf{q}_i)$ is the angular spectrum of the pump beam. Therefore,
\begin{eqnarray}
\label{coincidence}
C(\boldsymbol{\rho}_s, \boldsymbol{\rho}_i) 
&=& |\mathcal{C}_{2}|^{2} \, |\mathcal{W}(\mathbf{R}; Z_{0})|^{2}\,,\nonumber
\end{eqnarray}
where $\mathcal{W}(\mathbf{R}; Z_{0})$ is the transverse spatial profile of the pump beam propagated to $Z_{0}$ with
\begin{equation}
\label{Z0}
\frac{1}{Z_{0}} = \frac{\omega_{s}}{\omega_{p}} \frac{1}{z_{s}} + \frac{\omega_{i}}{\omega_{p}} \frac{1}{z_{i}}\,,
\end{equation}
\begin{equation}
\label{Xequation}
\mathbf{R} = \eta_s\,\boldsymbol{\rho}_{s} + \eta_i\,\boldsymbol{\rho}_{i}\,,
\end{equation}
where $\eta_j = \omega_j Z_0/(\omega_p z_j)$ ($j=s,i$) and $z_j$ denotes the distance from the crystal to the detection plane of photon $j\,$. Note from Eq.\eqref{Z0} that $\eta_s + \eta_i=1\,$. Thus, the pump intensity profile at \(z = Z_0\) is imprinted onto the coincidence rate through the coordinate \(\mathbf{R}\) defined in Eq.~\eqref{Xequation}.
%

%
We next consider two special cases of interest. First, we consider the situation where one detector, let us say the idler's,
is fixed at $\boldsymbol{\rho}_i=\mathbf{0}$ and the other (signal) is scanned along the line $y_s=0\,$. 
The corresponding coincidence count rate becomes
\begin{equation}
C(x_s,0) =
|\mathcal{C}_{2}|^{2} \, |\mathcal{W}(\eta_s x_s; Z{_0})|^{2}.
\label{1Dcc}
\end{equation}
This configuration shows that the full spatial structure of the pump beam can be observed in the signal photon at its respective scale.
%

Another interesting case arises when we emulate a two-photon detector by scanning the signal and idler detectors simultaneously, keeping both at the same transverse position, \(\boldsymbol{\rho}_i = \boldsymbol{\rho}_s\). This corresponds to the situation in which the signal and idler photons originate from the same point in the transverse plane. For simplicity, we consider that both detectors are scanned along the line $y_s=y_i=0$ and $x_s=x_i\equiv x\,$. In this case, the coincidence rate distribution is given by
\begin{equation}
C(x,x) =
|\mathcal{C}_{2}|^{2} \, |\mathcal{W}(x; Z_{0})|^{2}\,.
\label{1Dc0}
\end{equation}
Consequently, the two-photon coincidence distribution follows that of the pump on a spatial scale narrower than the pattern obtained by fixing one detector.

\label{sec:exp_setup}

\textit{Experimental setup.} -- The experimental setup is sketched in Fig.~\ref{fig:exp_setup}. A CW laser at 405~nm is set to horizontal polarization by a half-wave plate (HWP) before being sent to a spatial light modulator (SLM), which allows us to prepare arbitrary spatial modes. A $4f$ system images the SLM plane onto a 10~mm periodically poled potassium titanyl phosphate (PPKTP) type-0 crystal placed in a temperature-stabilized oven. The beam waist at the crystal plane is $w_0 = 0.1$~mm. A second HWP, positioned before the crystal, rotates the pump polarization to the vertical direction, as required by the phase-matching condition. The oven temperature is set to produce nondegenerate photon pairs at 780~nm (signal) and 840~nm (idler) via SPDC.

The photon pairs propagate freely over $z = 60$~cm to the detection system, where a dichroic mirror separates the signal and idler beams into two distinct spatial paths. In each path, a narrow bandpass filter precedes the coupling into a multimode optical fiber, which guides the photon to a single-photon counting module (SPCM).

Both detectors are mounted on motorized translation stages, controlled individually, which enables two detection configurations of interest. In the first, only the signal-arm performs the spatial scan in the transverse plane, while the idler detector remains fixed at the origin. This yields the conditional coincidence distribution $C(\eta_s x_s,0)\,$, allowing us to characterize the heralded single-photon pattern. In the second, the signal and idler fiber couplers, as well as the dichroic mirror, are scanned simultaneously as a rigid unit across the transverse plane, collecting light from a single transverse coordinate, even though one collects signal photons and the other idler photons. This yields the joint coincidence distribution $C(x,x)\,$, granting access to the two-photon transverse spatial state, i.e., the biphoton state. In both cases, photon coincidences are recorded by appropriate electronics.

\begin{figure}
    \centering
    \includegraphics[width=1.0
    \linewidth]{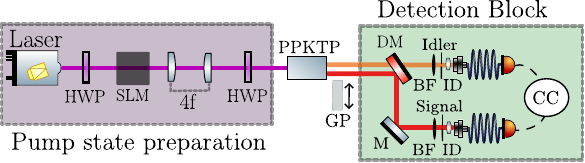}
    \caption{Experimental setup. A 405~nm CW laser is shaped by a spatial light modulator (SLM) and imaged onto a nonlinear crystal, generating photon pairs at 780~nm (signal) and 840~nm (idler) via SPDC. A dichroic mirror (DM) separates signal and idler photons and sends them to single-photon detection in coincidence (CC). Two configurations are used: in the heralded one, only the signal coupler is scanned; in the biphoton one, both couplers are scanned synchronously. In the biphoton configuration a glass plate (GP) is inserted to demonstrate phase sensitivity improvement.
    }

    \label{fig:exp_setup}
\end{figure}
%
%
\label{sec:result}
\textit{Results.} -- The pump laser is structured as a self-splitting beam $\Psi(x,y,z)$ according to Eq.~\eqref{Psi}, where the relative phase $\theta_c$ between the HG components controls the transverse profile at $z=0$, as illustrated in Fig.~\ref{fig:superposition}(a). In Fig.~\ref{fig:scanxy_panels_abc}, we report the conditional coincidence distribution obtained by 2D scanning of the heralded single-photon in the far-field plane, with the idler detector kept fixed. Panels (a), (b), and (c) show the heralded-photon transverse pattern for $\theta_c = 0, \pi/2$, and $\pi$, respectively. These results confirm that the three-dimensional spatial structure imprinted on the pump at $z=0$ is faithfully transferred to the quantum domain: the heralded single-photon follows the expected pattern illustrated at $z = 5z_R$ in Fig.~\ref{fig:superposition}(b) for the respective values of $\theta_c$.

\begin{figure}
    \centering
    \includegraphics[width=0.9\linewidth]{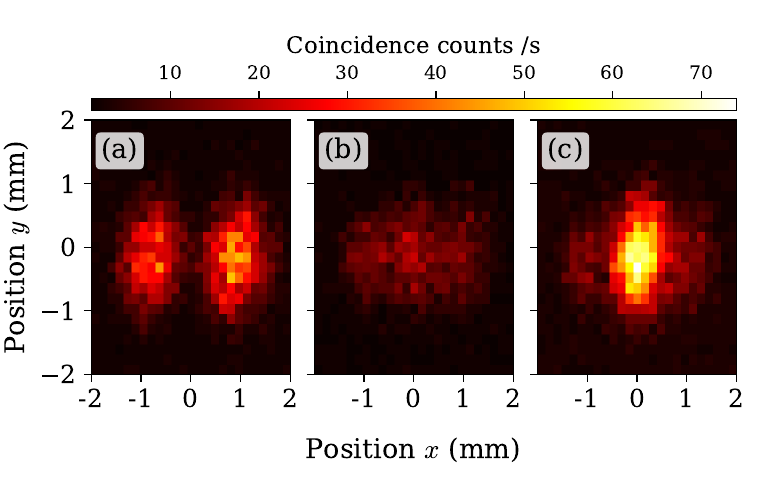}
    \caption{Experimental characterization of the self-splitting quantum spatial 
correlation at the heralded single-photon level. The pump beam is prepared 
as the self-splitting superposition $\Psi$ at $\tilde{z}=0$. Panels (a), (b), and 
(c) show the coincidence counts resolved over the transverse scan position 
$(x, y)$ in the detection plane located at $\tilde{z} = 5$ for pump relative 
phases $\theta_c = 0$, $\pi/2$, and $\pi$, respectively.}
    \label{fig:scanxy_panels_abc}
\end{figure}

As shown above, tuning the relative phase $\theta_c$ of the self-splitting pump $\Psi$ provides control over the spatial quantum correlations at the detection plane, demonstrating quantum self-splitting at the level of the heralded single-photon. Since the robustness of quantum correlations against obstacles is a central issue in quantum communication~\cite{mclaren2014self,nape2018self}, we propose that structured light engineered through the Gouy phase can be used to circumvent opaque objects. Exploiting the fact that our spatial mode splits into two lobes, we investigated whether the quantum spatial correlation could survive the presence of a vertical obstacle placed along the path of the heralded single-photon. In Fig.~\ref{fig:scanx_obstacle}(a), we show the heralded single-photon transverse distribution along $x$ (with $y=0$) for two pump configurations: a Gaussian $\mathrm{HG}_{00}$ mode (purple squares) and the superposition $\Psi$ with $\theta_c = \pi$ at $z=0$ (blue circles). Figure~\ref{fig:scanx_obstacle}(b) shows the same measurements performed with a slit (1~cm long, 1.2~mm wide) inserted in the region where the self-splitting occurs. When the pump is Gaussian, the obstacle produces a clear depletion in the heralded single-photon pattern. In contrast, when the pump is prepared in $\Psi$, the quantum spatial correlation is essentially unaffected, passing around the obstacle intact. This demonstrates that Gouy-phase engineering constitutes a useful tool to preserve quantum correlations through opaque obstacles.

\begin{figure}
    \centering
    \includegraphics[width=0.9\linewidth]{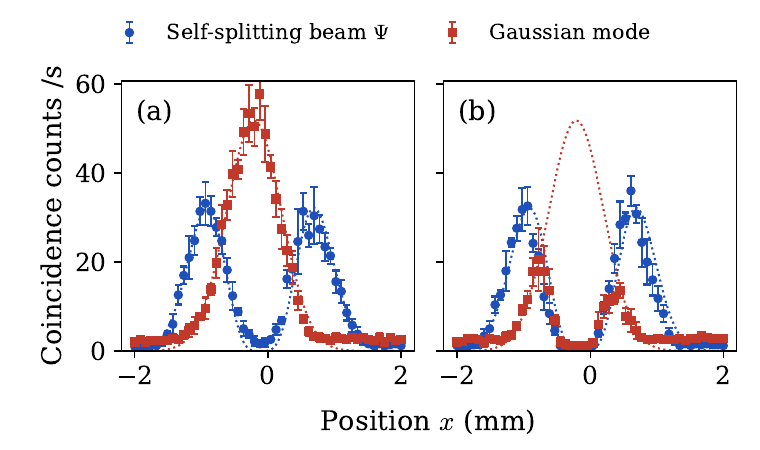}
    \caption{Coincidence counts as a function of transverse $x$ position. (a) Without obstacle: the $HG_{00}$ mode (purple squares) fitted by a Gaussian profile (dotted curve), and the superposition state (blue circles) fitted with the dashed curve. (b) With an obstacle ($\approx 1.2$ mm slit), the same theoretical profiles in (a) are overlaid as a reference. }
    \label{fig:scanx_obstacle}
\end{figure}

\begin{figure}[b]
    \centering
    \includegraphics[width=0.47\textwidth]{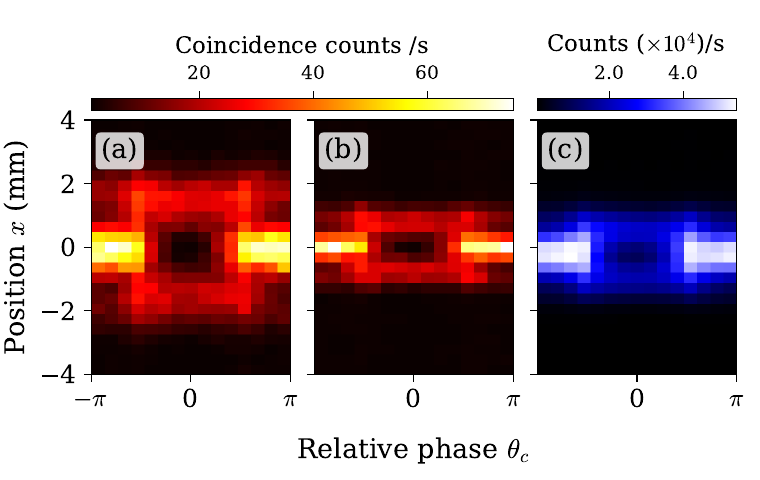}
    \caption{Experimental emulation of the self-splitting effect by tuning 
$\theta_c$. Varying $\theta_c$ reproduces the effective 
longitudinal evolution along the optical axis discussed in the simulation 
of Fig.~\ref{fig:superposition}. Panels (a) and (b) show the coincidence 
counts from an $xz$-emulated scan of the signal arm for two detection 
configurations: (a) the heralded single-photon, with the idler fixed at 
$x_i = 0$, and (b) the biphoton state, obtained by jointly scanning both 
detectors with $x_s = x_i$. Panel (c) shows the single-photon counts 
measured with an attenuated pump under the same scan conditions as (a).}
    \label{fig:scan2d-spdc}
\end{figure}

Figure~\ref{fig:scan2d-spdc} shows the transverse spatial profile along $x$ as the relative phase $\theta_c$ is varied, which emulates the propagation distance $z$ illustrated in the simulation of Fig.~\ref{fig:superposition}. Panel~(a) displays the heralded single-photon distribution, panel~(b) the two-photon joint distribution, and panel~(c) the pump intensity distribution. In (a), the idler detector is kept fixed at the origin (peak coincidence) while the signal detector is scanned; in (b), both detectors are scanned together, so that they collect light from the same point---strictly, a small area---in the transverse plane; in (c), a single detector records the pump intensity profile. All three distributions follow the same self-splitting dynamics inherited from the pump. The single-photon distribution in (a) is wider than the others, since it evolves with the signal wavelength, whereas the two-photon distribution in (b) propagates according to its de~Broglie wavelength~\cite{fonseca99b}, which coincides with that of the pump in (c). These measurements demonstrate the transfer of the angular spectrum from the pump to the two-photon conditional distributions, confirming the generation of both single-photon and two-photon self-splitting. 

In Fig.~\ref{fig:scan1d_minus}(a), we show the biphoton distribution obtained by scanning both detectors jointly, so that they probe the same point in the transverse plane. Black squares correspond to $\theta_c = \pi$ and blue circles to $\theta_c = 0$. For $\theta_c = \pi$, the coincidences display a single peak at the origin. In contrast, for $\theta_c = 0$, two symmetric coincidence peaks emerge, and each down-converted pair populates either the right or the left peak, never splitting between them. This is confirmed in panel (b), where the two detectors are displaced in opposite directions from the origin: the probability of finding the two photons on opposite sides ($x_i = - x_s$) vanishes, consistent with photon bunching within each peak. We interpret this configuration as spatial access to a NOON-like state, generated purely through the spatial quantum correlations of the structured pump. Conversely, for $\theta_c = \pi$ in the counter-scan, coincidences reappear at symmetric positions away from the origin, consistent with a spatially antibunched two-photon structure.

\begin{figure}
    \centering
    \includegraphics[width=0.9\linewidth]{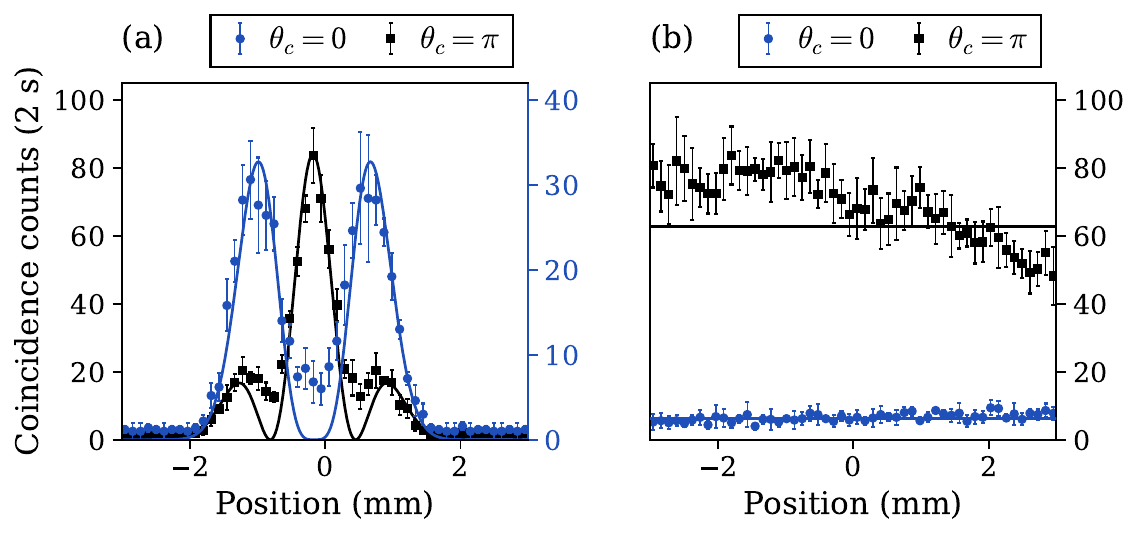}
    \caption{Self-splitting biphoton distribution for $\theta_c = 0$ (blue circles) 
and $\theta_c = \pi$ (black squares). Panel (a) shows the standard 
joint-scan configuration ($x_i = x_s$), while panel (b) shows the 
counter-scan configuration ($x_i = -x_s$).}
    \label{fig:scan1d_minus}
\end{figure}

We also observe in Fig.~\ref{fig:scan2d-spdc}(b) that the biphoton state exhibits a length scale nearly half of that of the single-photon distribution in Fig.~\ref{fig:scan2d-spdc}(a), in agreement with the two-photon de~Broglie wavelength. Moreover, the longitudinal evolution of the self-splitting structure is effectively equivalent to that of a Mach--Zehnder interferometer. To support this interpretation and to demonstrate that such a Mach--Zehnder-like interferometer can be exploited for phase-sensing applications, we measured the phase shift induced by a glass plate inserted in one of the peaks (arms) of the spatial interferometer, as indicated in Fig.~\ref{fig:exp_setup}. 
Figure~\ref{fig:scanx_glassplate}(a) displays the far-field interference pattern of the pump beam (405~nm) after a glass plate is introduced in its left near-field peak, simulating a phase difference in one arm of the Mach--Zehnder-like interferometer. As expected, the central peaks shift in position, revealing the phase delay induced by the glass plate. Panel (b) shows the spatial coincidence distribution for the down-converted photons under two detection configurations: with the idler fixed at $x_i = 0$ and with joint detection ($x_i = x_s$). Notably, the fringe spacing in the joint-detection case is smaller, equal to the pump wavelength, which demonstrates the super-resolution arising from the effective NOON state.

\begin{figure}
    \centering
    \includegraphics[width=0.9\linewidth]{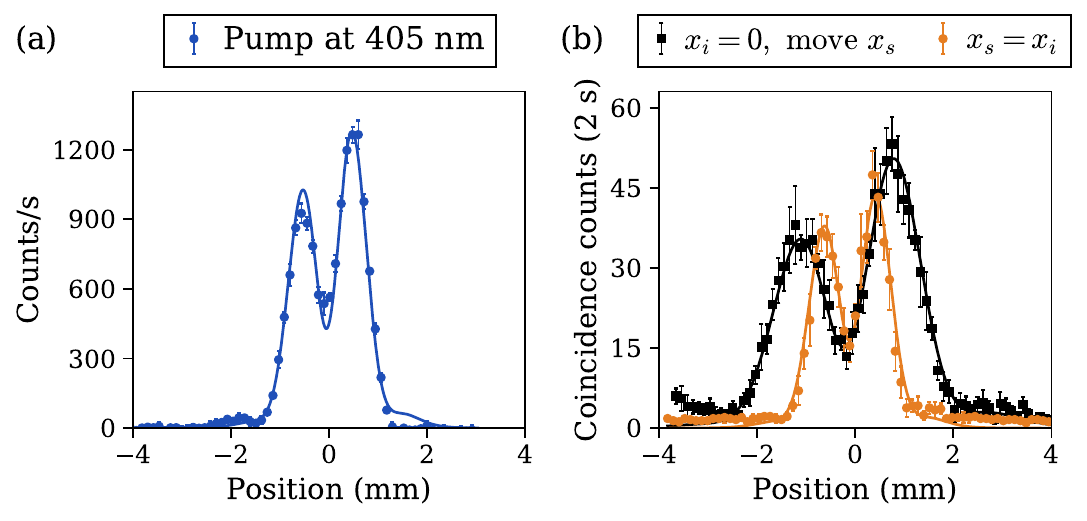}
    \caption{Phase-sensing in the self-splitting interferometer. 
(a)~Far-field interference of the pump beam (405~nm) after inserting 
a glass plate in the left near-field peak. 
(b)~Spatial coincidence distribution of the down-converted photons for 
two detection configurations: idler fixed at $x_i = 0$ (heralded 
single-photon) and joint detection ($x_i = x_s$).
    }\label{fig:scanx_glassplate}
\end{figure}


\textit{Conclusion.} -- Our results establish Gouy phase engineering as a powerful tool for controlling the propagation of quantum correlated photon pairs. By properly structuring the pump beam of SPDC, we demonstrated that quantum correlations can undergo self-splitting and recombination, implementing a Mach–Zehnder-like interferometer within the two-photon probability distribution. The observation of heralded single-photon and two-photon NOON state interferences highlights the potential of this approach for quantum metrology and quantum information processing. These findings open new avenues for exploiting spatial phase control to engineer complex quantum states of light and to design novel interferometric schemes at the single- and multi-photon level.

	\section*{Acknowledgments}

This work has been supported by the following Brazilian research agencies: Conselho Nacional de Desenvolvimento Cient\'{\i}fico e Tecnol\'ogico (CNPq - DOI 501100003593), Coordena\c c\~{a}o de Aperfei\c coamento de Pessoal de N\'\i vel Superior (CAPES DOI 501100002322), Fundação Carlos Chagas Filho de Amparo à Pesquisa do Estado do Rio de Janeiro (FAPERJ), Funda\c c\~{a}o de Amparo \`{a} Pesquisa do Estado de Santa Catarina (FAPESC - DOI 501100005667, DOI 2025TR001683), Instituto Nacional de Ci\^encia e Tecnologia INCT-IQNano 406636/2022-2 and INCT-DQ 408783/2024-9), and Fundação de Amparo à Pesquisa do Estado de São Paulo (FAPESP, Grants No. 2021/06823-5, No. 2022/15036-0, and No. 2022/15035-3).

 \bibliographystyle{apsrev4-2}
 \bibliography{biblio}
\end{document}